\documentstyle[preprint,aps]{revtex}

\begin{document}
\draft

\title{Benchmark calculations for polarization observables in 3N scattering} 

\author{A. Kievsky and M. Viviani}
\address{ Istituto Nazionale di Fisica Nucleare, Piazza Torricelli 2,
          56100 Pisa, Italy }
\author{S. Rosati}
\address{ Dipartimento di Fisica, Universita' di Pisa, Piazza Torricelli 2,
          56100 Pisa, Italy }
\author{D. H\"uber}
\address{Los Alamos National Laboratory, M.S. B283, Los Alamos, NM 87545, USA}
\author{W. Gl\"ockle and H. Kamada}
\address{Institut f\"ur Theoretische Physik II,
 Ruhr-Universit\"at Bochum, D-44780 Bochum, Germany}
\author{H. Wita\l a and J. Golak}
\address{Institute of Physics, Jagellonian University, PL-30059 Cracow,
Poland}

\maketitle

\abstract{High precision benchmark calculations for phase-shifts and
mixing parameters as well as observables 
in elastic neutron-deuteron scattering below the
deuteron breakup threshold are presented using a realistic nucleon-nucleon
potential. Two totally different methods, one using a variational principle
in configuration space and the other solving the Faddeev equations
in momentum space are used and compared to each other. The agreement achieved
in phase-shifts and mixing parameters as well as in the polarization
observables is excellent. The extreme sensitivity of the vector analyzing power 
A$_y$ to small changes of the phase shifts and mixing parameters 
is pointed out.} 

\bigskip

\noindent PACS numbers 21.45.+v, 25.10.+s, 03.65.Nk

\bigskip

\noindent key words: three-nucleon system, N-d scattering, polarization
  observables, phase-shift and mixing parameters, S-matrix.

\bigskip

\noindent corresponding author:

\noindent Alejandro Kievsky, Dipartimento di Fisica,
 Via Buonarroti 2, 56100 Pisa - ITALY \\
tel: +39-50-844546, Fax: +39-50-844538, e-mail: kievsky@pisa.infn.it

\newpage

\section{Introduction}

  A complete theoretical description of the three-nucleon (3N) system 
  is still limited by our knowledge of the nuclear interaction. Recently,
  progress has been achieved by optimally tuning various NN potential models
  to the NN data base, which lead to a fit with a $\chi^2$ per datum
  very close to 1. Even by using these modern NN potentials in triton 
  calculations \cite{hueber97}
  the well known underbinding problem is still present. 
The calculated binding energies lie between 7.6 - 8.0 MeV. 
A possible way to overcome this difficulty consists in including three-nucleon 
interaction (TNI) terms in the 3N Hamiltonian, usually fitted to reproduce the 
correct experimental binding energy  of 8.48 MeV. There are
  various models for TNI, arising from the $\pi$-$\pi$ exchange \cite{TNI}, 
  exchanges of heavier mesons\cite{TNI2} or having
  more phenomenological forms\cite{Urbana}. 
The investigation of the TNI effects must not be limited to
the 3N bound state but should be extended to the
  3N continuum. A prerequisite to that is a well grounded theoretical 
approach and the numerical control of its application to
3N scattering problems. Under this respect, much progress has been 
made in recent years \cite{pisa}\cite{witala}\cite{bochum}\cite{report}. 
As shown in \cite{report}, the overall agreement to measured 3N observables
 using modern NN potentials is quite good, but there are exceptions. Among 
 them we can recall the 3N nucleon vector analyzing power $A_y$, 
which depends very sensitively
 on the $^3P_j$ NN force components \cite{Ay1}, or the deuteron vector 
analyzing power
 $iT_{11}$, which shows a similar sensitivity. 
Both these observables have specific dependencies in terms of 3N S-matrix 
elements:
they are determined mainly by the $^4P_J$-parameters \cite{hueber95a,KRTV96}.
Such strong dependencies require very accurate calculations. 
The aim of the present article is to demonstrate that extremely accurate 
numerical results can be achieved.
 These two 3N observables $A_y$ and $iT_{11}$
 are of special interest, since present theoretical descriptions are 
about 30\% off the experimental data in the low energy region and
up to now no explanation has been found for the discrepancy 
\cite{Ay1,Ay2,Ay4,Ay3}.

 One way to parametrize the amplitude for elastic Nd scattering is in terms
 of the partial-wave decomposed S-matrix elements 
$S^J_{\lambda' \Sigma' \lambda\Sigma}$. 
Here $J$ is the total 3N angular 
momentum,
 $\lambda$ and $\lambda'$ the orbital angular momenta of a nucleon in relation
 to the deuteron and $\Sigma$ and $\Sigma'$ the total spins of 
 the deuteron and the third nucleon. The S-matrix elements can be 
expressed in terms of phase-shift
 and mixing parameters. As already stated above, the analyzing powers $A_y$ 
and $iT_{11}$ show extreme dependencies on some of them. As it will be
discussed in the next section,
differences of about 1\% in some phase-shift parameters can lead to
 differences in these observables as large as 10\% and more.
 Other observables, as the tensor analyzing powers and the spin-transfer and
 spin-correlation coefficients, are sensitive to states with high values
 of $\lambda$, typically $\lambda\ge 2$, which are also important when
 phase-shift analysis (PSA) are performed.

 In the present work we provide benchmark calculations
 for 3N scattering observables as well as for S-matrix parameters
 below the deuteron breakup threshold using
 a realistic NN interaction. Two different techniques are used to calculate
 the S-matrix elements. The Bochum-Cracow group solves the Faddeev equations in
 momentum space as described in \cite {witala} and \cite{report}. 
 The Pisa group uses the Kohn variational
 principle in configuration space \cite{pisa}\cite{KVR93}\cite{K97}. 
 Both techniques have been used in
 \cite{hueber95} but limiting the comparison just to  
 phase-shift and mixing parameters
 for states with total angular momentum $J\le 7/2$.
 Here we extend the investigation to 
 a number of observables by taking into account also 
 states with higher $J$ values which are 
 needed for a complete convergence of all considered observables. 
 At  the same time we increase the accuracy in order to demonstrate the
 numerical reliability of both methods to an unprecedented degree.
 Special emphasis is laid to the numerical accurate description of the 
 vector analyzing power $A_y$, since it poses a severe theoretical
 puzzle.

 The results obtained by the two techniques for the
phase shift and mixing parameters are 
presented in the next section. Those for the observables are
reported in section 3. The conclusions are the
content of the final section.

\section{Phase shift and mixing parameters} 

The two approaches used for numerical applications in this article
have been described previously. The Pisa group uses 
the Pair correlated Hyperspherical Harmonic (PHH) basis to expand the
scattering wave function \cite{pisa} and the corresponding $S$-matrix
is obtained using the complex form of the Kohn variational principle
\cite{K97}. The Bochum-Cracow group solves the Faddeev equations
 for the breakup operator in momentum space \cite{witala,report}.
 The complex transition matrix  for elastic scattering is then   gained
  by quadrature.

The comparison between the results of the two techniques has been 
performed using one of the Argonne NN potentials, namely 
the AV14\cite{AV14} one, which has all
the complexities of a modern NN interaction built in. 
 Our choice has been motivated by the 
fact that  this potential was used in many benchmark
calculations in the past, especially in \cite{hueber95}.
The incident nucleon
 laboratory energy has been fixed at $E_{lab}=3.0$ MeV, 
just below the deuteron
 breakup threshold. 

 In the correlated hyperspherical method the  pattern
 of convergence for the S-matrix has been studied including
in the n-d wave functions channels  with 
 increasing angular momenta values. 
Let us denote by $\ell_\alpha$ and $L_\alpha$ the orbital angular
 momentum quantum numbers associated to the Jacobi vectors
of the 3N system in the channel $\alpha$, and define
$K_0 = l_\alpha + L_\alpha$. The choices $K_0 \le$ 2(3), 4(5), 6(7) mean
 that all channels with $l_\alpha + L_\alpha \le$ 2(3), 4(5), 6(7)
 have been included  for positive (negative) parity states. Of course,
the number of hyperspherical components for each channel has been increased 
until convergence has been reached. As an example,
the numbers of the channels pertaining to the choices  $K_0 = 2, 4, 6$, 
for the $J^\Pi = 1/2^+$ state, result to be $N_c = 10, 18, 26$, respectively. 
 Channels with higher $K_0$ values were
checked to give completely disregardable contributions.

 In the Faddeev method in momentum space the
 convergence for the S-matrix is studied increasing the total two-body
 angular momentum $j_{max}$ up to which the NN force is taken into account.
 We went up to $j_{max} =8$ for $J=1/2$ and 3/2. 
It turned out that with $j_{max}=6$ a complete
convergence for the phase-shifts and mixing parameters was achieved. So,
$j_{max}=6$ was used in the calculations for higher $J$ values.
As an example, the values $j_{max} = 2$, 4, 6, 8 for $J = 1/2$ correspond to
 18, 34, 50, 66 channels, respectively. For higher $J$'s the number of channels
increases up to 34, 98, 194 and 322, respectively.

The results for the phase-shift and mixing parameters are displayed in
Table \ref{tab02} for states up to $J=9/2$. The Pisa numbers are 
calculated with $K_0=6(7)$ for the positive (negative) parity states
and the Bochum-Cracow numbers with $j_{max}=6$. 
For states with higher $J$ values the nuclear exchange term ($PG_0^{-1}$ as
given in eq.~(2.14) of ref.~\cite{hueber95}) 
is sufficient, as has been standard use
in the Bochum-Cracow approach. In the notation of the Pisa group this amounts
to the following easy manner to compute the S-matrix. 
In the states with high relative 
angular momentum $\lambda$ the incident nucleon and the deuteron
are well separated due to centrifugal barrier effects.
Therefore, the asymptotic form of the wave function
gives essentially a correct description of the system. 
The reactance matrix ($K$-matrix) of the system is
given in this approximation (symmetrized Born approximation) by

\begin{equation}
\label{Born1}
  K^J_{\lambda'\Sigma'\lambda\Sigma}= 
  \sum_{ij} <\lambda'\Sigma' ,i;J|H-E|\lambda\Sigma ,j;J>
\end{equation}
where the ket $|\lambda\Sigma ,i,J>$ describes an incident nucleon $i$ and
a deuteron in a relative angular 
momentum $\lambda$ state. $\Sigma$
is the total spin of the deuteron and the nucleon. 
Using the fact that the asymptotic state is solution of the
free hamiltonian plus the interaction between particles $j,k$, the following
simpler form is obtained

\begin{equation}
\label{Born2}
  K^J_{\lambda'\Sigma'\lambda\Sigma}= 
  3 \sum_i\sum_{j\ne 1} <\lambda'\Sigma' ,i;J|V(2,3)|\lambda\Sigma ,j;J> \ .
\end{equation}
A further approximation consists in
retaining the most important term (Born approximation):
\begin{equation}
\label{Born3}
  K^J_{\lambda'\Sigma'\lambda\Sigma}= 
  3 \sum_{j\ne 1}<\lambda'\Sigma' ,1;J|V(2,3)|\lambda\Sigma ,j;J> \ .
\end{equation}
This form is equivalent to the nuclear exchange term used by the 
Bochum-Cracow group:
\begin{equation}
\label{Born4}
  K^J_{\lambda'\Sigma'\lambda\Sigma}= 
  3 <\lambda'\Sigma' ,1;J|PG_0^{-1}|\lambda\Sigma ,1;J> \ .
\end{equation}
The $S$-matrix is simply obtained using the relation 
$S=(1+iK)(1-iK)^{-1}$.
 For the low energy used in
this study, the Born
approximation has been used for states with angular momenta $J\ge 11/2$.
The results obtained in this approximation for the phase-shift 
and mixing parameters 
by the two groups completely overlap and
are given in table \ref{tabBorn} for states 
from $J=11/2$ up to $J=15/2$. As it will be shown in the next section, 
higher $J$ states give disregardable contributions to the observables
at the energy considered here.

The numbers presented in Table \ref{tab02}, obtained by the two
different methods, agree between each other to within less than 0.1\%. 
There are a few exceptions, where the differences go up to 0.7\%.
 This clearly demonstrates the power and reliability
 of both methods, which are totally different also under the 
respect of the adopted numerical procedures.
 The comparison of our new results to the older ones in Table II of 
Ref.~\cite{hueber95}
gives a clear idea of the improvements in the numerical accuracy. 
The phases given in Table II of \cite{hueber95}
by the two methods show some disagreements 
in the 3rd digit and, sometimes, even in the 2nd digit.
Now we have much better agreement. Only very seldom there is disagreement
in the 3rd digit of a mixing parameter. So, we have now an improved 
agreement in the phase-shift parameters
of one more digit with respect to \cite{hueber95}.

A direct comparison of the $S$-matrix represents a severe 
numerical test for both methods.
The $S$-matrix is part of the wave function and its elements are
very sensitive to the different contributions of the potential. This is
put in evidence in Table \ref{tab02} where some parameters
converged only after the inclusion of high components in the expansion. To
construct the subtle details of the wave function is always a
difficult task. To this end, extended and denser grids of points have been
used in the numerical solution of the equations.
Stable numerical results have been obtained using
the integral  equations  for calculating the $S$-matrix (Bochum group) and
the Kohn second order estimate (Pisa group). Fortunately, 
as shown in the next section, for the observables there is not
such a strong dependence on the details of the wave function and a number
of grid points like those ones used in previous works is adequate.

\section{3N elastic scattering observables}

 We begin by showing in Fig.~\ref{comp} 
the perfect agreement among the observables
 evaluated by the two methods. The two results for the observables are
represented by the solid and long-dashed lines in Fig.~\ref{comp}, which
are always indistinguishable from each other. 
These calculations are based on the phases
 of Table I evaluated with highest accuracy.

 However we can relax those high requests and still keep a very good agreement
 among the predictions of the two methods. For the Bochum-Cracow calculations
 we can lower the two-nucleon angular momentum $j_{max}$ up to which we keep
 the NN force different from zero from $j_{max}=6$ to $j_{max}=3$ and still
 describe the observables within about 1 percent. Some examples are shown 
in Fig.~\ref{jmax} calculated with $j_{max}=2$,
4, and 6, respectively. The lines for $j_{max}=4$ and 6 are indistinguishable,
whereas the lines for $j_{max}=2$ exhibits small deviations from the other
lines for some of the observables.
The changes in the phases gained by going to $j_{max}$ higher than 3 
cancel out in the observables 
or occur for small phase-shifts and mixing 
parameters which are not so 
important for the determination of the observables. 
The same phenomenon can be seen by
increasing the numerical accuracy. Though a lot of phase-shifts and mixing
parameters show sensitivity to numerical details, the observables do not.
For the Pisa approach most of the large phase-shifts are already converged 
for $K_0=4(5)$ (see Table~\ref{tab02}), whereas the smallest observables
converge only for $K_0=6(7)$, as shown if Fig.~\ref{fig:zzz}.

Next let us regard the convergence of the observables with respect to 
$J$ and $\lambda$ separately. Since it does not matter which of the
two sets of phases we are using, we carry through that study with the
Bochum-Cracow phases.  

Let us start with the convergence in $J$. It turns out that the convergence
in $J$ can be described in a very systematic way. For spin observables whose
magnitude is about 0.1, or greater than that, the convergence is
reached already for $J\le 11/2$. In Fig.~\ref{Jcon} we show as a
typical example $K_y^{y'}$. For the next class of observables with
magnitudes about 10 times  smaller convergence is found for $J\le
13/2$. Here we have chosen as a typical example $A_y$ also shown in
Fig.~\ref{Jcon}. For the next class, again by about 
a factor of 10 smaller, even $J\le 15/2$ is not quite sufficient, as can be
seen in the example $T$ in Fig.~\ref{Jcon}, and states up to $J=25/2$
have to be taken into account. Thus the convergence in $J$ is
strongly correlated to the magnitudes of the observables.

Let us now examine the convergence with respect to $\lambda$.
For a given $J$ and parity $\Pi$, the S-matrix elements
are all coupled to each other. Therefore, 
limiting the calculation of the observables by a maximal
$\lambda$ (instead of a maximal $J$) means that for some $J^\Pi$
states, only parts of the corresponding S-matrices are 
taken into account. However,  this procedure
does no harm and one gets always a very nice convergence. 
Actually, at these low energies, the
contributions of the waves with large values of $\lambda$ are 
suppressed by  the centrifugal barrier, and therefore the convergence
in $\lambda$ is usually faster than that in $J$.

As an example we show in Fig.~\ref{lacon}  the $K_y^{z'z'}$ and $T$
observables. As it can be seen in the figure, for $\lambda\le 5$ $T$
is already converged, and $\lambda\le 4$ is not too bad, either. As one
can see for example from Table \ref{tab02}, $\lambda\le 5$ means that the  
highest $J$ taken into account is 13/2. On the other hand the much larger
observable $K_y^{z'z'}$ is fully converged only for
$\lambda\le 6$, which means one needs phases up to $J=15/2$. In $J$ this
observable reaches convergence earlier, for $J\le 13/2$.
But this is the exception. In most cases the convergence in
$\lambda$ is faster than in $J$.

Now let us demonstrate the extreme  sensitivity of $A_y$ with respect
to tiny changes of some phase shift parameters, namely the $^4P_J$ 
phase-shifts and the $\epsilon^{3/2-}$ mixing parameter.
We modified them individually by 1\%. The effect on $A_y$ is shown in Table
\ref{tabay} (see also Table 2 in Ref.~\cite{hueber95a}). 
Clearly the sensitivity of $A_y$ to the $^4P_J$ phase-shifts
is quite dramatic - an enlargement factor of up to nearly 20 from changes in 
the phases to changes in the observable are found. There are no such
extremely strong sensitivities to the NN $^3P_j$ phase-shifts \cite{hueber95a},
although they alone determine $A_y$. In \cite{hueber95a} the biggest
enlargement factor was reported for $^3P_0$ to be 3.5. In view of that extreme
sensitivity it is interesting to see which $A_y$ would result by using the
phases of Ref.~\cite{hueber95}, which were calculated not with such an 
extreme accuracy  as in this article.  
Thereby it is interesting to note how
the more accurate calculations in this study change all of these
four parameters which dominate $A_y$ 
for both approaches compared to the older and less
accurate numbers in \cite{hueber95}: the changes are 0.24\% (0.20\%) for
$^4P_{1/2}$, 0.42\% (0.42\%) for $^4P_{3/2}$, 0.30\% (0.65\%) for $^4P_{5/2}$
and 0.55\% (1.65\%) for $\epsilon^{3/2-}$ in the Bochum-Cracow (Pisa) 
case. Now the $A_y$
resulting from the phases of \cite{hueber95}
is shown in Fig.~\ref{comp} (beside other observables) 
in comparison to the present 
best result. We see a small shift for the Bochum-Cracow result and a
larger one for the Pisa result. This can be illustrated further by assuming 
that $A_y$ changes linearly with changes of the phases around their 
present values. Using Table~\ref{tabay}
together with the small changes of the present phases to the ones of
Ref.~\cite{hueber95} one indeed finds that $A_y$ should change by about 10\% 
(1\%) for the Pisa (Bochum) case. In other words in one case we see
a stability for the resulting observable, in the other case not. The
simple reason is that the expansion in the PHH components in \cite{hueber95} 
was
truncated in a nonuniform manner for the different states and in 
\cite{hueber95} it
was not foreseen that even those small changes in the phases would effect
certain observables in a magnified manner. On the other hand if a consistent
treatment of all states is performed the individual changes of the phases
are smoothed out in the observables, as it was the case for the Bochum-Cracow
calculation in Ref.~\cite{hueber95}. 
In the Pisa group papers successive to Ref.~\cite{hueber95}
and in the present paper the PHH expansion has been consistently
carried through for all states with a correct calculation of $A_y$.

\section{Summary and Conclusions}

In the present paper  benchmark calculations for phase-shift
and mixing parameters, as well as for observables in elastic neutron-deuteron
scattering below the deuteron breakup threshold are presented. We used
the realistic AV14
 NN potential.
Two {\it ab initio} completely different methods have been used to calculate
the quantities of interest. 
The approach of the Pisa group is based on the correlated hyperspherical 
expansion of the wave function in configuration space and uses the 
complex Kohn variational principle to determine the S- or the K-matrix. 
The approach of the Bochum-Cracow group is based on an exact 
technique for solving the Faddeev equations in momentum space. 

The results obtained for the phase-shift and mixing parameters by
means of the two approaches show nearly perfect agreement. Also the
calculated observables agree very well with each other. 
This demonstrates that both the  variational approach of the Pisa
group and the integral equation method of the Bochum-Cracow group are  
equally  well suited for high
accuracy calculations of the elastic nd
scattering observables below the breakup threshold.

The comparison with the older results of Ref.~\cite{hueber95} shows that 
the numerical accuracy in the phase-shift and mixing 
parameters has increased by one more digit. 
The changes in the phase-shifts had only very small effects on the
results for the observables of the Bochum-Cracow group. The  situation
was different for the Pisa results, since in Ref.~\cite{hueber95}  
the phases were evaluated with different accuracy requirements
for the different states and an observable like $A_y$ which exhibits
extreme sensitivity does not tolerate that.
This shows that it is important to construct
the observables  by using $S$--matrix elements calculated at the same
approximation level. In this case, in fact, a more rapid convergence
with respect to the number of terms included in the internal structure
of the n-d states and a lower sensitivity to the numerical accuracy is
achieved for the observables. 
In fact, the $S$--matrix elements are part of the wave
function and therefore rather sensitive to the subtle aspect of the
structure of the state. On the contrary, the observables are average
quantities where the small details  of the wave function
are somewhat smeared out.

Also the convergence of the observables along the total
three-body angular momentum $J$ and the relative angular momentum
$\lambda$ has been studied. We found that the convergence of the
observables with the total three-body angular momentum $J$ is strongly
correlated to the magnitude of the  observables. 
Though the convergence of the observables with the angular momentum
$\lambda$ is in most cases faster than in $J$, it is less
systematic and has therefore to be checked with more care.
Therefore a PSA has to treat the more phase-shifts as free parameters the
smaller the considered observables are.

In the present paper, attention has been only paid to a realistic two nucleon
interaction without the inclusion of TNI terms. This will be the object
of a future investigation.

In conclusion, n-d scattering states at energies below the
deuteron breakup threshold  can be constructed  equally accurately
by the two methods presented.
It is grateful that the phase-shift and mixing parameters
can be calculated within a precision of about  $0.1\%$, comparable to
the one obtainable for bound states.

\begin{acknowledgements}
This work was supported in part by the
Deutsche Forschungsgemeinschaft under Project No. Hu 746/1-3 (D.H.)
and Project No. Gl 87/24-1 (H.W.)
and was performed in part under the auspices of the U.S.
Department of Energy. The numerical calculations of the Bochum-Cracow group
have been performed 
on the Cray T90 of the H\"ochstleistungsrechenzentrum in J\"ulich, 
Germany. 
\end{acknowledgements}

\newpage

\begin{table}
\begin{tabular} {c|c||c|c|c||c|c|c||}
&&\multicolumn{3}{c||}{Bochum}&\multicolumn{3}{c||}{Pisa}\\
\hline
$J^{\Pi}$&$\delta_{\Sigma\lambda}$&$j_{max}=2$&$j_{max}=4$&
$j_{max}=6$&$K_0=2(3)$&$K_0=4(5)$&$K_0=6(7)$\\
\hline
    &$\delta_{(3/2)2}$&-3.897&-3.903&-3.904&-3.899&-3.905&-3.905\\
${1\over2}^{+}$				   
    &$\delta_{(1/2)0}$&-35.35&-34.84&-34.81&-35.33&-34.81&-34.81\\
    &$\eta$           & 1.179& 1.247& 1.251& 1.271& 1.252& 1.253\\
\hline                                     
    &$\delta_{(1/2)1}$&-7.479&-7.524&-7.529&-7.534&-7.533&-7.533\\
${1\over2}^{-}$				   
    &$\delta_{(3/2)1}$& 25.10& 25.06& 25.06& 25.04& 25.05& 25.05\\
    &$\epsilon$       & 7.268& 7.253& 7.254& 7.252& 7.255& 7.255\\
\hline                                     
    &$\delta_{(3/2)0}$&-70.47&-70.48&-70.48&-70.52&-70.50&-70.50\\
    &$\delta_{(1/2)2}$& 2.439& 2.422& 2.421& 2.421& 2.420& 2.420\\
${3\over2}^{+}$				   
    &$\delta_{(3/2)2}$&-4.204&-4.214&-4.215&-4.216&-4.216&-4.216\\
    &$\eta$           &-.3963&-.3889&-.3881&-.3869&-.3873&-.3874\\
    &$\epsilon$       & .7745& .7766& .7785& .7747& .7801& .7800\\
    &$\xi$            & 1.451& 1.438& 1.438& 1.429& 1.438& 1.438\\
\hline                                     
    &$\delta_{(3/2)3}$& .9466& .9443& .9441& .9425& .9436& .9436\\
    &$\delta_{(1/2)1}$&-7.145&-7.186&-7.191&-7.201&-7.195&-7.195\\
${3\over2}^{-}$				   
    &$\delta_{(3/2)1}$& 26.44& 26.42& 26.41& 26.39& 26.40& 26.41\\
    &$\eta$           &-3.854&-3.813&-3.809&-3.819&-3.806&-3.805\\
    &$\epsilon$       &-2.751&-2.764&-2.765&-2.762&-2.768&-2.765\\
    &$\xi$            &-.2400&-.2567&-.2574&-.2577&-.2573&-.2575\\
\end{tabular}
\pagebreak
\begin{tabular} {c|c||c|c|c||c|c|c||}
&&\multicolumn{3}{c||}{Bochum}&\multicolumn{3}{c||}{Pisa}\\
\hline                                           
$J^{\Pi}$&$\delta_{\Sigma\lambda}$&$j_{max}=2$&$j_{max}=4$&
$j_{max}=6$&$K_0=2(3)$&$K_0=4(5)$&$K_0=6(7)$\\
\hline
    &$\delta_{(3/2)4}$&-.2105&-.2109&-.2110&-.2113&-.2112&-.2111\\
    &$\delta_{(1/2)2}$& 2.401& 2.388& 2.386& 2.382& 2.385& 2.385\\
${5\over2}^{+}$				    
    &$\delta_{(3/2)2}$&-4.558&-4.567&-4.567&-4.571&-4.569&-4.569\\
    &$\eta$           &-2.084&-2.146&-2.152&-2.167&-2.157&-2.157\\
    &$\epsilon$       &-.3450&-.3266&-.3264&-.3387&-.3275&-.3280\\
    &$\xi$            &-.7637&-.7379&-.7356&-.7343&-.7365&-.7363\\
\hline                                          
    &$\delta_{(3/2)1}$& 26.40& 26.39& 26.38& 26.32& 26.35& 26.37\\
    &$\delta_{(1/2)3}$&-.4723&-.4760&-.4765&-.4771&-.4768&-.4767\\
${5\over2}^{-}$				    
    &$\delta_{(3/2)3}$& .9757& .9720& .9716& .9694& .9711& .9712\\
    &$\eta$           &-.3475&-.3596&-.3605&-.3593&-.3609&-.3609\\
    &$\epsilon$       & .5007& .5165& .5168& .5188& .5165& .5165\\
    &$\xi$            & .9566& .9832& .9844& .9943& .9847& .9845\\
\hline                                          
    &$\delta_{(3/2)2}$&-4.140&-4.142&-4.143&-4.151&-4.145&-4.144\\
    &$\delta_{(1/2)4}$& .1107& .1104& .1103& .1100& .1103& .1102\\
${7\over2}^{+}$				    
    &$\delta_{(3/2)4}$&-.2200&-.2204&-.2205&-.2208&-.2207&-.2209\\
    &$\eta$           &-.5130&-.4923&-.4905&-.4868&-.4894&-.4895\\
    &$\epsilon$       & .3575& .3694& .3683& .3695& .3680& .3686\\
    &$\xi$            & 1.266& 1.225& 1.221& 1.219& 1.222& 1.222\\
\end{tabular}
\pagebreak
\begin{tabular} {c|c||c|c|c||c|c|c||}
&&\multicolumn{3}{c||}{Bochum}&\multicolumn{3}{c||}{Pisa}\\
\hline
$J^{\Pi}$&$\delta_{\Sigma\lambda}$&$j_{max}=2$&$j_{max}=4$&
$j_{max}=6$&$K_0=2(3)$&$K_0=4(5)$&$K_0=6(7)$\\
\hline
    &$\delta_{(3/2)5}$&.04950&.04947&.04946&.04944&.04944&.04944\\
    &$\delta_{(1/2)3}$&-.4654&-.4683&-.4688&-.4695&-.4690&-.4689\\
${7\over2}^{-}$		
    &$\delta_{(3/2)3}$& 1.030& 1.027& 1.026&1.024&1.026&1.026\\
    &$\eta$           &-2.334&-2.311&-2.304&-2.298&-2.301&-2.308\\
    &$\epsilon$       &-.2120&-.2505&-.2527&-.2500&-.2524&-.2510\\
    &$\xi$            &-.7720&-.7790&-.7823&-.7842&-.7826&-.7873\\
\hline					   
    &$\delta_{(3/2)6}$&-.01170&-.01170&-.01170&-.01170&-.01170&-.01170\\
    &$\delta_{(1/2)4}$& .1088& .1086& .1085&.1082&.1084&.1084\\
${9\over2}^{+}$				   
    &$\delta_{(3/2)4}$&-.2292&-.2293&-.2294&-.2297&-.2297&-.2297\\
    &$\eta$           &-2.214&-2.218&-2.223&-2.231&-2.227&-2.226\\
    &$\epsilon$       &-.2049&-.1983&-.1954&-.1947&-.1897&-.1954\\
    &$\xi$            &-.8291&-.8289&-.8262&-.8251&-.8252&-.8259\\
\hline					   
    &$\delta_{(3/2)3}$& .9441& .9440& .9439&.9413&.9435&.9435\\
    &$\delta_{(1/2)5}$&-.02550&-.02550&-.02552&-.02553&-.02553&-.02553\\
${9\over2}^{-}$	
    &$\delta_{(3/2)5}$&.05151&.05150&.05149&.05147&.05147&.05147\\
    &$\eta$           &-.4781&-.4841&-.4863&-.4887&-.4867&-.4866\\
    &$\epsilon$       & .3316& .3268& .3277&.3273&.3275&.3276\\
    &$\xi$            & 1.146& 1.156& 1.161&1.164&1.162&1.162\\
\end{tabular}
\caption{\label{tab02} phase shifts and mixing parameters in terms of the
quantum numbers $j_{max}$ and $K_0$. The numbers in parenthesis for
$K_0$ refer to odd parity states. 
\\
}%{\bf Table 2}
\end{table}

\newpage

%\begin{table}[h]
\begin{table}
%\begin{center} 
\begin{tabular} {c|c|c|c|c|c}
$J^{\Pi}$&$\delta_{\Sigma\lambda}$& Bochum-Cracow-Pisa
&$J^{\Pi}$&$\delta_{\Sigma\lambda}$& Bochum-Cracow-Pisa \\
\hline
    &$\delta_{(3/2)4}$&-.2135 &&$\delta_{(3/2)7}$& .00280 \\
    &$\delta_{(1/2)6}$& .00603&&$\delta_{(1/2)5}$&-.02513 \\
${11\over2}^{+}$&$\delta_{(3/2)6}$&-.01216&
${11\over2}^{-}$&$\delta_{(3/2)5}$& .05311 \\
    &$\eta$           &-.4942 &&$\eta$           &-2.204 \\
    &$\epsilon$       & .2978 &&$\epsilon$       &-.1785 \\
    &$\xi$            & 1.142 &&$\xi$            &-.8481 \\
\hline		       
    &$\delta_{(3/2)8}$&-.00067 &&$\delta_{(3/2)5}$& .04968\\
    &$\delta_{(1/2)6}$& .00593 &&$\delta_{(1/2)7}$&-.00144\\
${13\over2}^{+}$&$\delta_{(3/2)6}$&-.01249&
${13\over2}^{-}$&$\delta_{(3/2)7}$& .00290\\
    &$\eta$           &-2.185 &&$\eta$           &-.4975 \\
    &$\epsilon$       &-.1660 &&$\epsilon$       & .2779 \\
    &$\xi$            &-.8640 &&$\xi$            & 1.125 \\
\hline
    &$\delta_{(3/2)6}$&-.01173 &&$\delta_{(3/2)9}$& .00016\\
    &$\delta_{(1/2)8}$& .00035 &&$\delta_{(1/2)7}$&-.00141\\
${15\over2}^{+}$&$\delta_{(3/2)8}$&-.00070&
${15\over2}^{-}$&$\delta_{(3/2)7}$& .00297\\
    &$\eta$           &-.4998 &&$\eta$           &-2.170 \\
    &$\epsilon$       & .2634 &&$\epsilon$       &-.1570 \\
    &$\xi$            & 1.111 &&$\xi$            &-.8752 \\
\hline		       
\end{tabular}
\caption{\label{tabBorn} Phase shifts and mixing parameters from
$J=11/2$ to $J=15/2$ in Born approximation.
}%{\bf Table 2}
%\end{center} 
\end{table}

%\begin{table}[h]
\begin{table}
%\begin{center} 
\begin{tabular} {ccr}
&$A_y|_{max}$&\%\\
\hline
AV14&0.3306\\
$^4P_{1/2}*1.01$&0.2960&$-$11.7\\
$^4P_{3/2}*1.01$&0.3083&$-$7.2\\
$^4P_{5/2}*1.01$&0.3362&18.4\\
$\epsilon^{3/2-}*1.01$&0.3362&1.7
\end{tabular}
\caption{\label{tabay} The effect of 1\% changes in the phases to which $A_y$
is most sensitive. Given is the value of $A_y$ in its maximum as well as the
change in percent in the maximum.
}%{\bf Table 2}
%\end{center} 
\end{table}

\newpage

\begin{figure}
\caption{\label{comp} The differential cross section $d\sigma /d\Omega$,
the tensor spin-correlation coefficient $T$,
the nucleon-to-nucleon vector spin-transfer coefficient $K^{y'}_y$,
the nucleon-to-deuteron tensor spin-transfer coefficient $K^{z'z'}_y$,
and the nucleon and deuteron vector analyzing powers $A_y$ and $iT_{11}$
at $E_{lab}=3$ MeV. The solid lines  are obtained from 
the 
Bochum-Cracow phases of this work, 
with the potential switched on up to $J=9/2$ and the nucleon
exchange term taken  into account from $J=11/2$ till 15/2. The long-dashed
lines are the same but obtained from the Pisa phases. The solid and long-dashed
lines are always indistinguishable from each other.
The short-dashed line is obtained for
$J^\Pi=1/2^\pm$, $3/2^\pm$, $5/2^\pm$, and $7/2^+$ from the Bochum-Cracow
phases of Ref.~\protect\cite{hueber95} and otherwise the phases from this 
paper.
The dotted lines are the same but obtained from the corresponding Pisa phases.}
\end{figure}

\begin{figure}
\caption[figure]{  \label{jmax} 
The differential cross section $d\sigma /d\Omega$, 
the nucleon vector analyzing power
$A_y$, the deuteron tensor analyzing power $T_{21}$, 
the nucleon-to-deuteron tensor spin-transfer
coefficient $K^{z'z'}_y$ and the two tensor spin-correlation coefficients
S and T, calculated by increasing the maximum allowed two-body angular momentum
$j_{max}$. Solid, long-dashed, and short-dashed lines refer to $j_{max}=6$,
4, and 2, respectively. Note that the solid and long-dashed lines do always
completely overlap.
}
\end{figure}

\begin{figure}[t]
\caption[figure]{  \label{fig:zzz} 
The differential cross section $d\sigma /d\Omega$, 
the nucleon vector analyzing power
$A_y$, the deuteron tensor analyzing power $T_{21}$, 
the nucleon-to-deuteron tensor spin-transfer
coefficient $K^{z'z'}_y$ and the two tensor spin-correlation coefficients
S and T, calculated by increasing the maximum allowed $K_0$ in the expansion
of the wave function.
Solid, long-dashed, and short-dashed lines correspond to maximum
$K_0=6(7),4(5),2(3)$.
}
\end{figure}

\begin{figure}
\caption{\label{Jcon} Convergence in $J$ for the nucleon-to-nucleon vector
spin-transfer coefficient $K_y^{y'}$, the nucleon
vector analyzing power $A_y$, and the tensor
spin-correlation coefficient $T$. Shown is the result obtained from
the phases with $J\le 25/2$ (solid line), $J\le 15/2$ (short-dashed
line), $J\le 13/2$ (dotted line), and  $J\le 11/2$ (long-short-dashed line).
}
\end{figure}

\begin{figure}
\caption{\label{lacon} Convergence in $\lambda$ for the tensor
spin-correlation coefficient $T$ and the
nu\-cle\-on-to-deu\-te\-ron tensor spin-transfer coefficient $K_y^{z'z'}$. 
Shown are the 
results obtained from the phases with $\lambda\le 6$ (long-dashed
line), $\lambda\le 5$ (short-dashed line), and $\lambda\le 4$ (dotted line).
The solid line is the same as in Fig.4}
\end{figure}

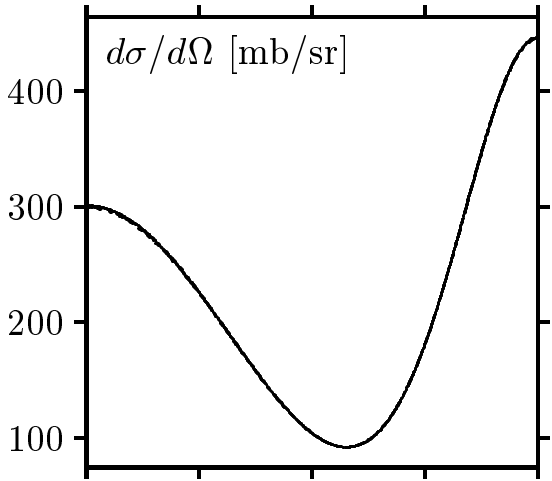
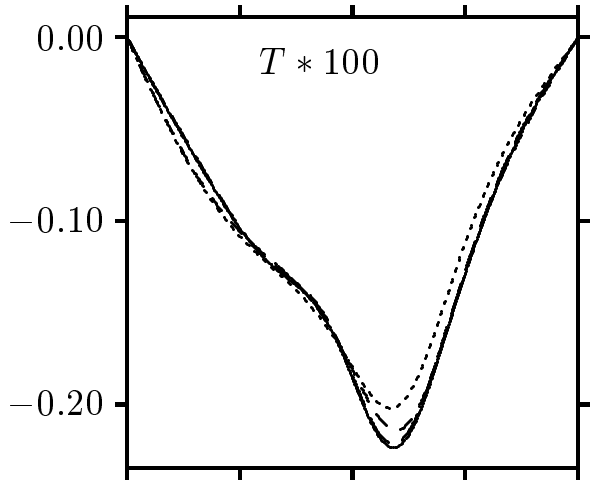
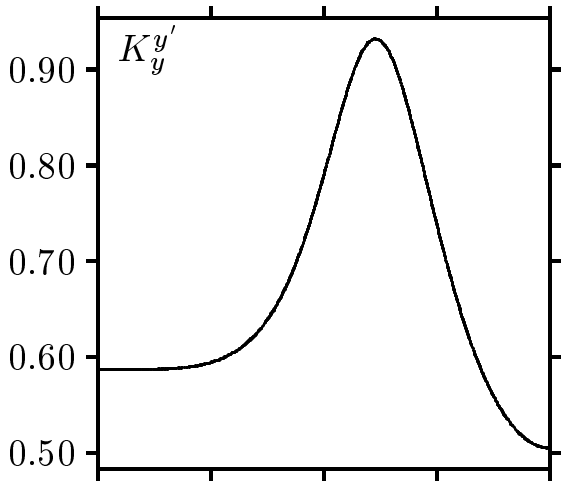
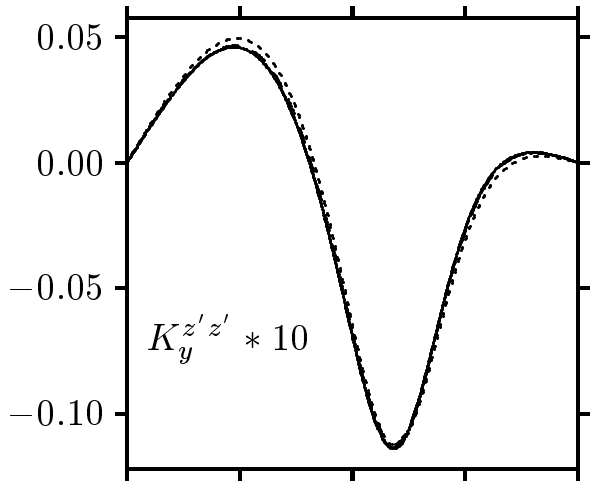
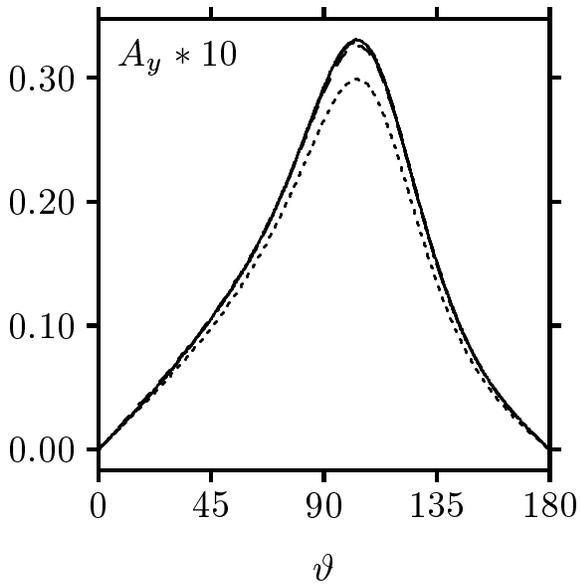
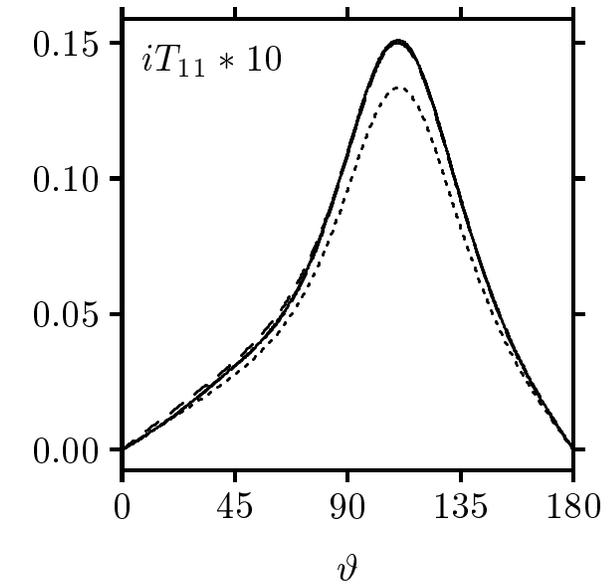
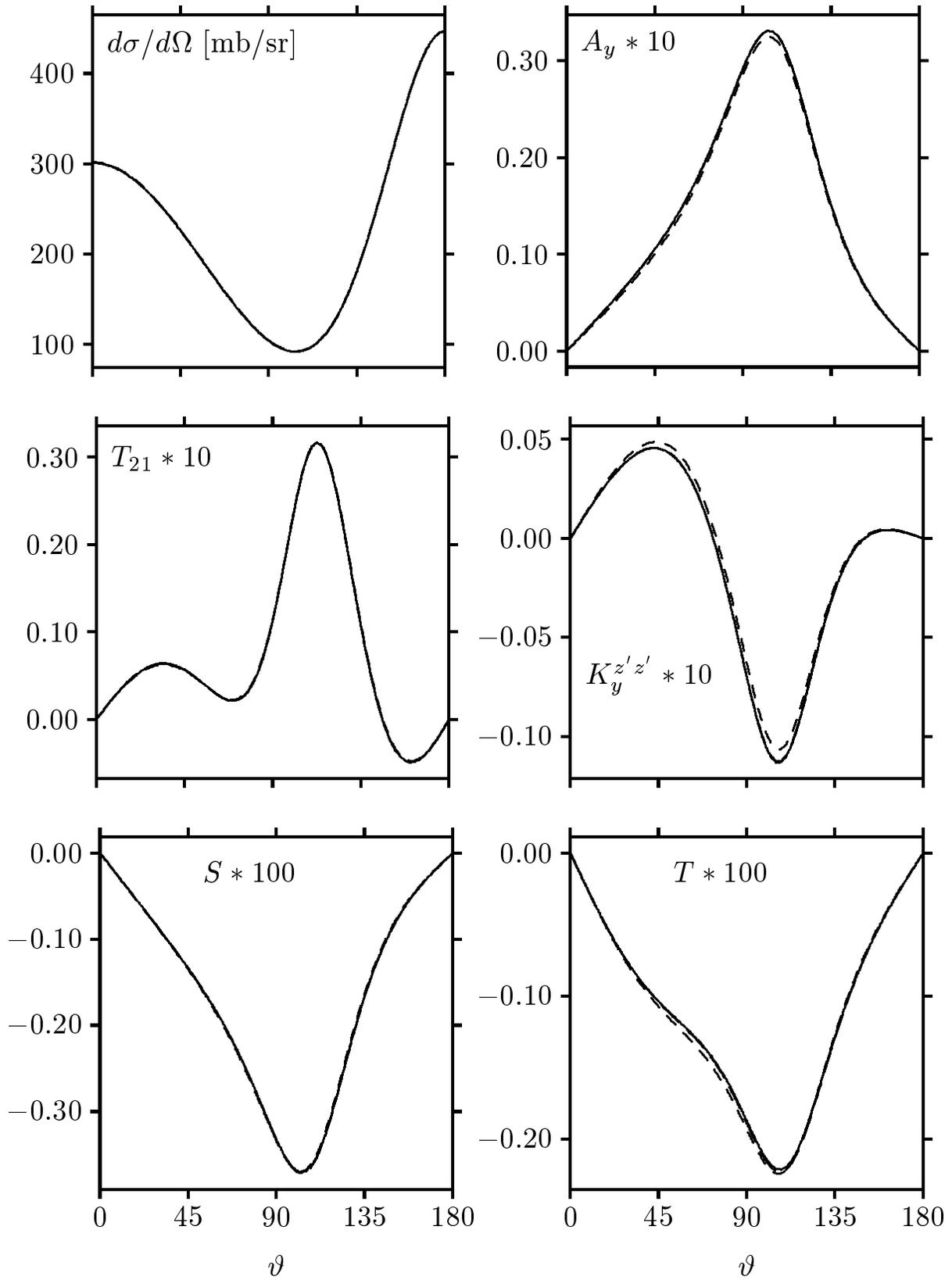
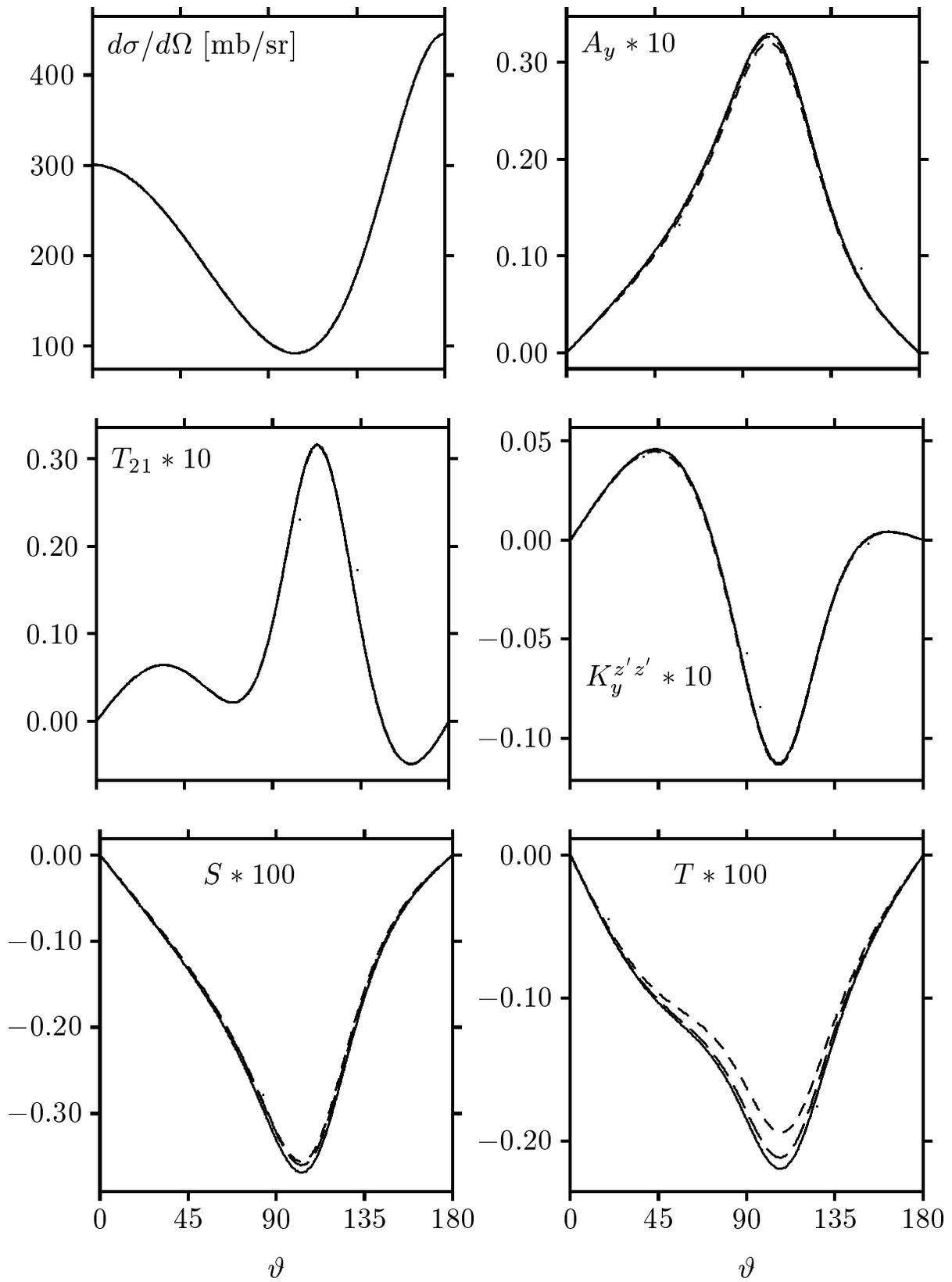
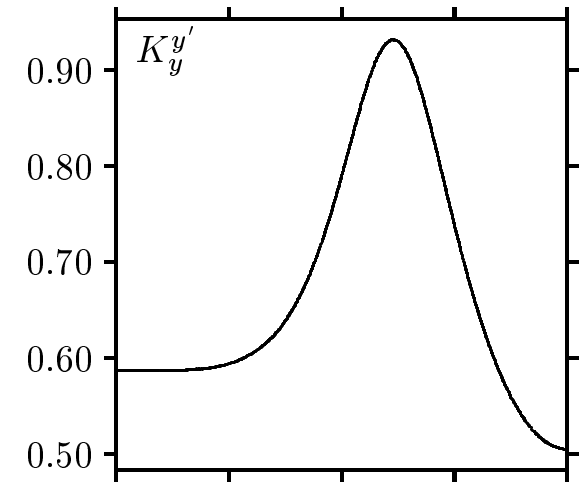
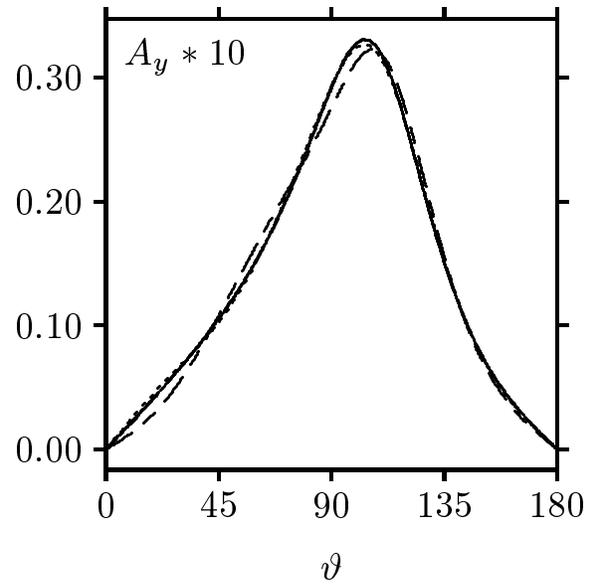
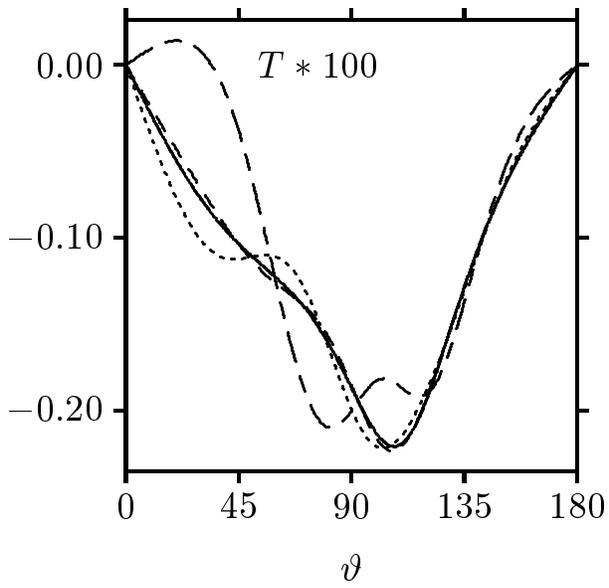
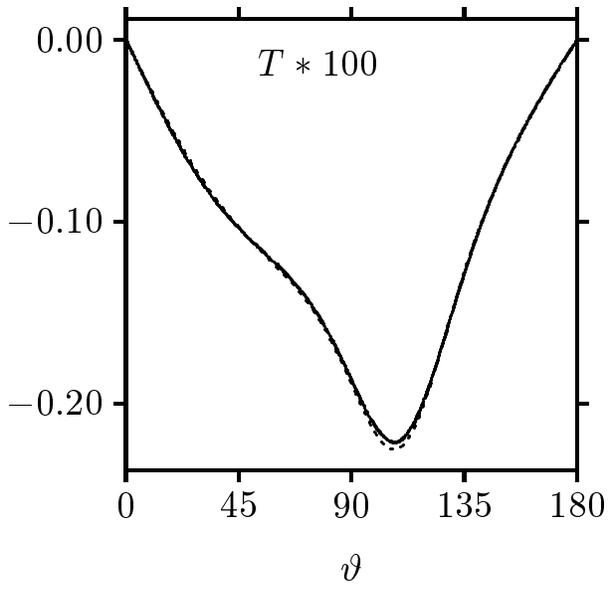
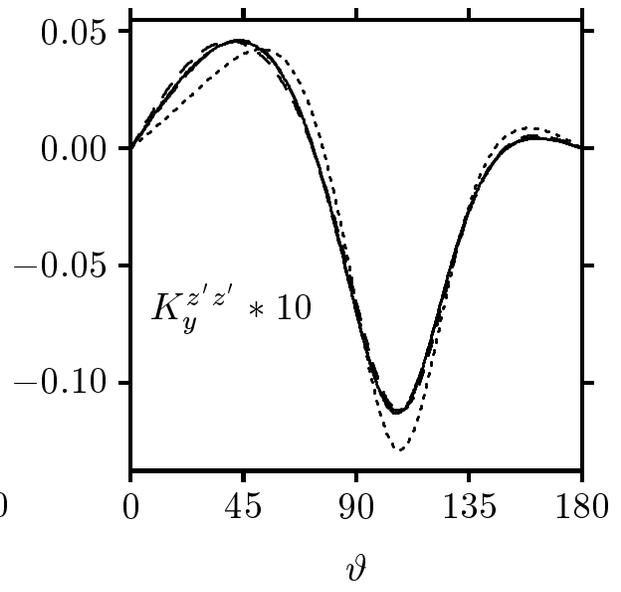
\end{document}